\documentclass[aps,prl,twocolumn,superscriptaddress,preprintnumbers]{revtex4-2} 

\usepackage[T1]{fontenc}
\usepackage[utf8]{inputenc}
\usepackage{lmodern}
\usepackage{amsmath,amssymb,amsthm,bm,booktabs}
\usepackage{hyperref}
\usepackage{graphicx} 
\usepackage{bm} 
\usepackage{times} 
\usepackage{xcolor}
\usepackage{physics}
\usepackage{multirow}
\usepackage{pifont}
\usepackage{braket}

\usepackage{tikz}

\definecolor{darkgreen}{HTML}{006400}

\newcommand{\F}{\mathbb F}
\newcommand{\E}{\mathbb E}
\newcommand{\Msch}{M_2^{\mathrm{Sch}}}
\newcommand{\Sch}{\mathrm{Sch}}

\begin{document}

\title{A Universal Budget for Entanglement and Nonlocal Non-Stabilizerness}
	
\author{Salvatore Marco Giampaolo}
\affiliation{Institut Ruđer Bošković, Bijenička cesta 54, 10000 Zagreb, Croatia}
	
\begin{abstract}
	Entanglement and nonlocal non-stabilizerness are distinct quantum resources encoded in the Schmidt spectrum of a bipartite pure state.
	For qubit systems, we establish a universal quadratic constraint on their simultaneous allocation, governed by the minimum number of logical qubits required on each side to encode the occupied Schmidt support.
	Additional spectral constraints reveal a strong asymmetry in their allocation with the Entanglement that can exhaust the logical capacity, whereas the maximal nonlocal non-stabilizerness grows only logarithmically with it.
	For a tensor-network cut crossed by a single separating virtual edge, the logical capacity is determined by the occupied bond dimension.
	We further derive a one-sided certificate for the resources of the untruncated state from a retained Schmidt spectrum and its discarded weight.
	For small discarded weight, the resulting corrections are linear in that weight.
	These results connect a universal resource budget to resource-aware control of tensor-network truncations.
\end{abstract}

\preprint{RBI-ThPhys-2026-34}

	
\maketitle

Entanglement is both a fundamental manifestation of quantum mechanics~\cite{NielsenChuang, Horodecki2009} and an organizing principle for quantum many-body systems~\cite{Osterloh2002, Vidal2003, Amico2008, Eisert2010, Laflorencie2016}.
It provides a resource for quantum communication~\cite{Bennett1993, Wehner2018}, computation~\cite{Jozsa2003, Vidal2003_simulation, Vidal2004}, and metrology~\cite{Giovannetti2006, Pezze2018}.
Its structure and scaling characterize phases and diagnose criticality~\cite{Vidal2003, Eisert2010}, while also constraining the efficiency of tensor-network representations~\cite{Verstraete2006, Schollwock2011, Haegeman2016, Paeckel2019, Hastings2007}.
Entanglement alone, however, neither exhausts quantum resources nor determines computational complexity.
Stabilizer states can be highly entangled yet efficiently tractable within the Clifford framework, while product states may carry extensive non-stabilizerness.
Because stabilizer states and Clifford operations alone are not universal, they must be supplemented by non-stabilizer resources to achieve universal quantum computation. 
This additional resource, known as non-stabilizerness, quantifies departure from the stabilizer sector~\cite{Gottesman1998, BravyiKitaev2005, Veitch2014, Howard2014, Leone2022}.
Beyond its computational role, non-stabilizerness has attracted growing interest in quantum many-body physics~\cite{Oliviero2022, Odavic2023, Catalano2026, Liu2026}, providing access to properties not captured by entanglement alone~\cite{Catalano2025, Li2026}.

For a bipartite state, non-stabilizerness may reflect both local properties and correlations across the partition.
The local-unitary orbit of a state is the set of states obtained by applying arbitrary unitary transformations independently on the two sides of the partition. 
It therefore represents all possible choices of local bases.
These transformations leave the Schmidt spectrum, and hence the entanglement, unchanged.
To isolate the irreducible nonlocal content, one minimizes a chosen non-stabilizerness measure over this orbit~\cite{Cao2025, Torre2026, Franchini2026, Sierant2026, Viscardi2026}.
The resulting resource is invariant under local unitaries, as is entanglement.
For pure states on a fixed bipartite Hilbert space, these orbits are completely characterized by the Schmidt spectrum, namely the nonzero eigenvalues of either reduced density matrix.
Entanglement and nonlocal non-stabilizerness are therefore distinct resources encoded in the same spectral data.
This shared origin raises a basic structural question: can they be allocated independently, or does the finite dimension of the occupied Schmidt support impose a common budget?

This structural question is directly relevant to tensor-network descriptions of many-body states.
Tensor networks represent many-body wave functions as contractions of tensors over virtual indices.
The dimensions of the corresponding virtual spaces, known as bond dimensions, constrain the correlations that can be represented and govern the computational cost~\cite{White1992, Schollwock2011, Orus2014}.
A joint resource bound tied to the finite Schmidt support would extend the standard bond-dimension constraint on entanglement to a simultaneous constraint on entanglement and nonlocal non-stabilizerness~\cite{Hastings2007, Schollwock2011}.
Since practical tensor-network calculations rely on Schmidt truncation, a controlled estimate would further connect the resources of the retained approximation to those of the underlying many-body state.

In this Letter, we establish such a budget for qubit systems.
For any pure state $\ket{\psi}$ of a qubit system and any bipartition, we prove
\begin{equation}
	\bigl[S_2(\psi)\bigr]^2 +\bigl[M_2^{\rm NLM}(\psi)\bigr]^2 \leq q^2 .
	\label{eq:main-bound}
\end{equation}
Here, $S_2$ is the R\'enyi-2 entanglement entropy.
The nonlocal non-stabilizerness $M_2^{\rm NLM}$ is defined as the minimum stabilizer R\'enyi-2 entropy over the local-unitary orbit associated with the bipartition~\cite{Cao2025, Torre2026}.
The \emph{logical Schmidt capacity} $q=\lceil\log_2 r_{\rm Sch}\rceil$ is the minimum number of logical qubits required on each side to encode the occupied Schmidt support of rank $r_{\rm Sch}$.
It is determined by the Schmidt spectrum and is therefore independent of unused dimensions in the physical subsystems.
Equation~\eqref{eq:main-bound} establishes a joint resource budget by limiting the combined amount of entanglement and nonlocal non-stabilizerness that can be supported within this common logical space.

For a canonical open-boundary MPS, the Schmidt rank across a cut equals the occupied bond dimension $\chi_o$. 
The corresponding capacity is therefore $q=\lceil\log_2\chi_o\rceil$.
For more general tensor-network geometries, a bipartition may cross several virtual bonds.
Their combined dimensions then provide the upper bound $q\leq\lceil\sum_{e\in\partial A}\log_2\chi_e\rceil$.
Equation~\eqref{eq:main-bound} thus supplements the standard entanglement bound $S_2\leq\log_2 \chi_o$~\cite{Schollwock2011, Orus2019, Cirac2021} with a joint constraint on the resources carried by a finite virtual space.
We also extend the result to Schmidt truncations of arbitrary retained rank. 
The retained spectrum and discarded weight then provide a certificate for the resources of the untruncated state.

The allocation of the two resources is strongly asymmetric. 
Entanglement can saturate the logical capacity, whereas nonlocal non-stabilizerness requires entanglement and its maximum possible value grows at most logarithmically with that capacity.
In terms of the occupied bond dimension, this maximal value scales as $O(\log\log\chi_o)$, in contrast to the possible $\log_2\chi_o$ scaling of entanglement.


To establish Eq.~\eqref{eq:main-bound}, we use an explicit representative of the local-unitary orbit, constructed from the ordered Schmidt spectrum~\cite{Cao2025, Torre2026, Franchini2026}.
For an arbitrary pure state $\ket{\psi}$ of a qubit system with bipartition $A|B$, the local-unitary orbit is uniquely characterized by the nonzero Schmidt spectrum $\boldsymbol{\lambda}= (\lambda_0,\ldots, \lambda_{r_{\rm Sch}-1})$, ordered nonincreasingly so that \mbox{$\lambda_x\geq\lambda_{x+1}$}.
Its rank $r_{\rm Sch}$ is the dimension of the occupied support, and the logical Schmidt capacity $q=\lceil\log_2r_{\rm Sch}\rceil$ specifies the minimum number of logical qubits required on each side to accommodate it.
Setting $D=2^q$ and padding the spectrum with zeros for $r_{\rm Sch}\leq x<D$, we define the canonically ordered Schmidt representative
\begin{equation}
	\ket{\psi_{\rm Sch}(\boldsymbol{\lambda})} =\sum_{x\in\mathbb F_2^q}\sqrt{\lambda_x}\, \ket{x}_{A_q} \! \ket{x}_{B_q},
	\label{eq:canonical-state}
\end{equation}
on two logical $q$-qubit registers $A_q$ and $B_q$.
Here, $\mathbb F_2=\{0,1\}$ and $x=(x_1,\ldots,x_q)\in\mathbb F_2^q$ labels their computational basis.
The added components carry zero Schmidt weight and leave the occupied support unchanged.
If the physical subsystems contain $n_A$ and $n_B$ qubits, respectively, the logical state in Eq.~\eqref{eq:canonical-state} is understood as embedded into the original Hilbert space by adjoining $n_A-q$ and $n_B-q$ local qubits in $\ket{0}$.
The resulting embedded state belongs to the local-unitary orbit of $\ket{\psi}$.
Since the added qubits are local product ancillas, they leave the Schmidt spectrum, and hence the entanglement, unchanged.
Moreover, the stabilizer R\'enyi-2 entropy is additive and vanishes on stabilizer ancillas.
The embedding therefore also leaves the stabilizer entropy of the canonical representative unchanged.

Since entanglement is constant along the local-unitary orbit, the Rényi-2 entanglement entropy can be written as $S_2(\psi)=S_2(\psi_{\rm Sch})=S_2(\boldsymbol{\lambda})=-\log_2P$, where $P=\sum_x\lambda_x^2$ is the purity of either reduced state.
On the other hand, the stabilizer R\'enyi-2 entropy of $\psi_{\rm Sch}$ is $M_2^{\rm Sch}(\boldsymbol{\lambda})=-\log_2\Phi_q$, where $\Phi_q=D^{-1}\sum_{u,k\in\mathbb F_2^q}A_{u,k}^4$ is the stabilizer purity and $A_{u,k}=\sum_{x\in\mathbb F_2^q}(-1)^{k\cdot  x} \sqrt{\lambda_x \lambda_{x\oplus u}}$~\cite{Torre2026, Franchini2026}.
The quantities $A_{u,k}$ are Walsh transforms of the pointwise products between the Schmidt-amplitude vector and its binary translations.
They encode the Pauli expectation values entering the stabilizer entropy of $\ket{\psi_{\rm Sch}}$, with $u$ specifying a binary translation and $k$ labeling the corresponding Walsh mode.
The dot product $k\cdot x=\sum_{j=1}^q k_jx_j$ is evaluated modulo two, while $x\oplus u$ denotes componentwise addition in $\mathbb F_2^q$.

These expressions show that $S_2(\boldsymbol{\lambda})$ and $M_2^{\rm Sch}(\boldsymbol{\lambda})$ depend on the same ordered Schmidt spectrum but probe it differently.
The former is determined by a positive second moment, without interference or cancellations.
The latter depends on an aggregate fourth moment of signed combinations of Schmidt amplitudes and is therefore sensitive to cancellations within the Walsh correlators and to constraints linking different sectors.
Using this spectral representation, we prove that every bipartite pure state of a qubit system satisfies
\begin{equation}
	\bigl[S_2(\boldsymbol{\lambda})\bigr]^2 +	\bigl[M_2^{\rm Sch}(\boldsymbol{\lambda})\bigr]^2
	\leq q^2 .
	\label{eq:canonical-bound}
\end{equation}
The proof, given in the Supplemental Material, combines analytical estimates with a recursion based on the canonical ordering of the Schmidt spectrum and rigorous interval-arithmetic certificates for the remaining finite cases.

Since the embedded canonical state is an admissible representative of the local-unitary orbit, its stabilizer entropy upper-bounds the optimized resource, \mbox{$M_2^{\rm NLM}(\psi)\leq M_2^{\rm Sch}(\boldsymbol{\lambda})$}.
Together with the local-unitary invariance of entanglement, Eq.~\eqref{eq:canonical-bound} therefore implies Eq.~\eqref{eq:main-bound}.
Although several partial results support the conjectured optimality of the canonical representative~\cite{Franchini2026, SierantSpectral2026, LiuCui2026}, whether it attains the global minimum over the local-unitary orbit remains unproved in general.
However, the resource-budget constraint does not rely on this conjecture.


Equation~\eqref{eq:main-bound} endows the logical Schmidt capacity with a resource-theoretic meaning by constraining the joint allocation of entanglement and nonlocal non-stabilizerness.
Combining this trade-off with additional spectral constraints reveals a pronounced asymmetry between the two resources.
A flat spectrum with $2^q$ nonzero eigenvalues, $\lambda_x=2^{-q}$, is the Schmidt spectrum of $q$ Bell pairs and is characterized by $(S_2,M_2^{\rm Sch})=(q,0)$.
Conversely, the independent relation $M_2^{\rm NLM}\leq M_2^{\rm Sch}\leq 2S_2$~\cite{Torre2026, Franchini2026} forces both non-stabilizerness measures to vanish in the absence of entanglement, ruling out the opposite allocation $(S_2,M_2^{\rm Sch})=(0,q)$ for $q>0$.
Hence, entanglement can exhaust the logical capacity without carrying nonlocal non-stabilizerness, whereas nonlocal non-stabilizerness cannot occupy the capacity on its own.
To quantify the stronger constraint obeyed by the canonical representative, let
$\mu(q)=\sup_{\boldsymbol{\lambda}:\lceil\log_2r_{\rm Sch}(\boldsymbol{\lambda})\rceil=q}M_2^{\rm Sch}(\boldsymbol{\lambda})$ denote its supremal value at capacity $q$.
Two complementary arguments bound $\mu(q)$. 
The first is geometric: Eq.~\eqref{eq:canonical-bound} confines the resource pair to a quarter disk, while $M_2^{\rm Sch}\leq2S_2$ confines it below a straight line. 
The largest value of $M_2^{\rm Sch}$ allowed by both constraints occurs where their boundaries meet, at $S_2=M_2^{\rm Sch}/2$. 
Substitution into Eq.~\eqref{eq:canonical-bound} gives $5(M_2^{\rm Sch})^2/4\leq q^2$ and hence $\mu(q)\leq2q/\sqrt{5}$. 

A stronger bound at large $q$ follows from the canonical ordering of the spectrum.
To prove it, we group the ordered eigenvalues into layers whose sizes double: the largest eigenvalue alone, then the next one, the next two, the next four, and so on.
Formally, these are the $q+1$ dyadic blocks $B_0=\{0\}$ and $B_j=\{2^{j-1},\ldots,2^j-1\}$ for $j=1,\ldots,q$. 
Each block has weight $W_j=\sum_{x\in B_j}\lambda_x$, and therefore $\sum_{j=0}^qW_j=1$.
The recursion derived in the Supplemental Material translates this ordering into the lower bound $\Phi_q\geq W_0^4 +14\sum_{j=1}^q W_j^4$; the factor $14$ counts the cross-block terms entering the recursion.
For nonnegative block weights satisfying $\sum_{j=0}^qW_j=1$, H\"older's inequality gives
$\Phi_q\geq(1+q/14^{1/3})^{-3}$.
Taking $-\log_2\Phi_q$ and retaining the stronger of this result and the geometric bound gives
\begin{equation}
	\mu(q)\leq \min\!\left\{ \frac{2q}{\sqrt{5}}, 3\log_2\! \left(1+\frac{q}{14^{1/3}}\right) \right\}.
	\label{eq:max-resource}
\end{equation}
The linear branch is tighter for $1\leq q\leq6$, whereas the logarithmic branch provides the stronger bound for all $q\geq7$.

The logarithmic scaling in Eq.~\eqref{eq:max-resource} naturally raises the question of whether it can be attained by a family of Schmidt spectra.
Ref.~\cite{LiuCui2026} shows that such behavior is realized by the marginal algebraic spectrum $\lambda_x=[(x+1)H_N]^{-1}$, with $x=0,\ldots,N-1$, $N=2^q$, and $H_N\!=\!\sum_{j=1}^N j^{-1}$.
For this family, $M_2^{\rm Sch}\!=\!3\log_2q\!+\!O(1)$.
These spectra are admissible in the supremum defining $\mu(q)$ and therefore provide a lower bound with the same leading term as the logarithmic branch of Eq.~\eqref{eq:max-resource}, whose asymptotic expansion is $3\log_2q-\log_2 14+o(1)$.
Combining the upper and lower asymptotics yields $\mu(q)=3\log_2q+O(1)$.
Thus, both the logarithmic dependence on capacity and its leading coefficient are optimal for the canonical Schmidt representative.
The uniform additive bound between the canonical and optimized resources established in Ref.~\cite{SierantSpectral2026} implies that the supremum of $M_2^{\rm NLM}$ at fixed capacity also scales as $3\log_2q+O(1)$.
These asymptotic results neither identify the maximizing spectra at finite $q$ nor determine the optimal additive terms.


The logical Schmidt capacity has a direct tensor-network interpretation through the virtual space actually occupied across a bipartition.
Consider an open-boundary MPS brought into mixed canonical form across a bond $\ell$~\cite{Schollwock2011, Orus2019, Cirac2021}.
The nonzero singular values carried by that bond are the square roots of the Schmidt eigenvalues, and their number defines the occupied bond dimension $\chi_o^{(\ell)}=r_{\rm Sch}^{(\ell)}$, namely the minimal bond dimension required for an exact representation across the cut.
Consequently, $q_\ell=\lceil\log_2\chi_o^{(\ell)}\rceil$, whereas the nominal bond dimension $\chi^{(\ell)}$ may include unused virtual directions and only satisfies $\chi_o^{(\ell)} \leq \chi^{(\ell)}$.

For this cut, the canonical resource budget in Eq.~\eqref{eq:canonical-bound} becomes $(S_{2,\ell})^2+(M_{2,\ell}^{\rm Sch})^2\leq q_\ell^2$. 
It supplements the usual entropy constraint $S_{2,\ell}\leq\log_2\chi_o^{(\ell)}\leq\log_2\chi^{(\ell)}$ by limiting how much Schmidt-representative non-stabilizerness can coexist with a given entanglement within the occupied virtual space.
Conversely, an exact representation of a prescribed resource pair requires $q_\ell\geq [(S_{2,\ell})^2+(M_{2,\ell}^{\rm Sch})^2]^{1/2}$, which can impose a larger minimum logical capacity than the entanglement-only condition $q_\ell\geq S_{2,\ell}$.
Equation~\eqref{eq:max-resource} further yields \mbox{$M_{2,\ell}^{\rm Sch}=O(\log\log\chi_o^{(\ell)})$} for large $\chi_o^{(\ell)}$ and, since $M_{2,\ell}^{\rm NLM}\leq M_{2,\ell}^{\rm Sch}$, the same asymptotic upper bound for $M_{2,\ell}^{\rm NLM}$.
The single-bond identification extends to any tensor network in which removing one virtual edge separates the two parts of the bipartition.
This includes the half-chain cut of an infinite MPS in Schmidt canonical form~\cite{Vidal2007} and any edge bipartition of a tree tensor network~\cite{Shi2006, Gerster2014}.
The relevant condition is therefore the presence of a single separating edge, rather than the spatial dimensionality of the physical system.

If the bipartition crosses a set $\partial A$ of virtual bonds, their dimensions instead provide the bounds $r_{\rm Sch}(A|B)\leq\prod_{e\in\partial A}\chi_e$ and $q(A|B)\leq\lceil\sum_{e\in\partial A}\log_2\chi_e\rceil$.
The product need not be fully occupied because distinct virtual-boundary configurations may map to linearly dependent physical states, and some may map to zero.
Defining the effective occupied boundary dimension as $\chi_{o,\partial A}=r_{\rm Sch}(A|B)$ restores the exact relation $q(A|B)=\lceil\log_2\chi_{o,\partial A}\rceil$.
By definition, $\chi_{o,\partial A}$ counts the independent Schmidt components across the full bipartition and need not equal the product of the individual bond dimensions.
Thus, the resource budget remains intrinsic to the state across the bipartition, while the network geometry determines how its logical capacity is encoded or bounded within a particular representation.


For a generic pure state of a many-qubit system, the Schmidt spectrum across a bipartition contains $2^L$ nonzero eigenvalues, where $L=\min\{|A|,|B|\}$ is the number of qubits in the smaller subsystem.
Although symmetries, constraints, or other special structures may reduce this number, the generic exponential scaling makes storing and manipulating the full spectrum impractical already for moderately large $L$.
Tensor-network calculations therefore typically truncate the spectrum to its $\chi$ largest eigenvalues.
Our resource bound applies directly to the normalized retained spectrum, while the discarded weight turns it into a rigorous one-sided certificate for the resources of the untruncated state.

Let $\lambda_0\geq\lambda_1\geq\cdots\geq\lambda_{r_{\rm Sch}-1}>0$ denote the ordered nonzero spectrum, and retain its first $\chi$ entries, with $1\leq\chi\leq r_{\rm Sch}$.
The discarded and retained weights are $\epsilon_\chi=\sum_{i\geq\chi}\lambda_i$ and $w_\chi=1-\epsilon_\chi>0$, respectively.
The normalized retained spectrum is $\widetilde{\lambda}_i=\lambda_i/w_\chi$ for $i<\chi$.
We denote its entanglement and Schmidt-representative non-stabilizerness by $\widetilde S_2$ and $\widetilde M_2^{\rm Sch}$, respectively.
Hence, setting $r=\lceil\log_2\chi\rceil$ and padding the retained spectrum with zeros when necessary, we have $\widetilde S_2^2+(\widetilde M_2^{\rm Sch})^2\leq r^2$.

Recall that, for the Schmidt spectrum padded to dimension $D=2^q$, the stabilizer purity is $\Phi_q=D^{-1}\sum_{u,k}A_{u,k}^4$, where $A_{u,k}=\sum_x(-1)^{k\cdot x}\sqrt{\lambda_x\lambda_{x\oplus u}}$.
Expanding $A_{u,k}^4$ and summing over $k$ rewrites $\Phi_q$ as a homogeneous polynomial of degree eight in the Schmidt amplitudes with nonnegative coefficients.
Setting the discarded amplitudes to zero can therefore only decrease $\Phi_q$.
The remaining unnormalized amplitude vector is $\sqrt{w_\chi}$ times the normalized retained one, so degree-eight homogeneity and invariance under zero padding give $\Phi_q\geq w_\chi^4\widetilde\Phi_r$, or equivalently $M_2^{\rm Sch}\leq\widetilde M_2^{\rm Sch}-4\log_2w_\chi$, where $\widetilde\Phi_r=2^{-\widetilde M_2^{\rm Sch}}$.
Zero padding removes any power-of-two restriction on the retained rank $\chi$.

Defining $a_\chi=-2\log_2w_\chi$, the normalized retained spectrum, padded with zeros, majorizes the full spectrum, which gives $S_2\ge\widetilde S_2$, whereas $P\ge w_\chi^2\widetilde P$ yields $S_2-\widetilde S_2\le a_\chi$. 
Hence, $0\le S_2-\widetilde S_2\le a_\chi$, while the preceding estimate gives $M_2^{\rm Sch}\le\widetilde M_2^{\rm Sch}+2a_\chi$.
Combining these bounds with the retained-spectrum budget and applying the triangle inequality yields the one-sided certificate
\begin{equation}
	\begin{aligned}
		\sqrt{[S_2(\boldsymbol{\lambda})]^2\!+\![M_2^{\rm Sch}(\boldsymbol{\lambda})]^2}
		&\!\leq\!
		\sqrt{(\widetilde S_2)^2\!+\!(\widetilde M_2^{\rm Sch})^2}
		\!+\!\sqrt{5}\,a_\chi\\
		&\!\leq r+\sqrt{5}\,a_\chi .
	\end{aligned}
	\label{eq:truncation-certificate}
\end{equation}
Since $M_2^{\rm NLM}(\psi)\le M_2^{\rm Sch}(\boldsymbol{\lambda})$, the same right-hand side bounds the budget for the fully optimized nonlocal resource.

The additive corrections to the two resources depend only on the discarded weight, with $a_\chi =2\epsilon_\chi/\ln 2+O(\epsilon_\chi^2)$.
They are therefore linear for small discarded weight, uniformly over normalized retained spectra and with no explicit dependence on the full Schmidt rank.

In MPS calculations, the retained Schmidt spectrum and discarded weight are available at each monitored Schmidt truncation.
Equation~\eqref{eq:truncation-certificate} gives the usual discarded-weight threshold a quantitative interpretation in terms of resource bounds.
For each bond, one may increase $\chi_\ell$ until $\sqrt{5}\,a_{\chi,\ell}$ falls below a prescribed tolerance on the one-sided resource correction.
The retained resource norm $\widetilde B_\ell=[(\widetilde S_{2,\ell})^2+(\widetilde M_{2,\ell}^{\rm Sch})^2]^{1/2}$ can be evaluated directly from the Schmidt spectrum, without local-unitary optimization.
For $r_\ell=\lceil\log_2\chi_\ell\rceil>0$, the ratio $\eta_\ell=\widetilde B_\ell/r_\ell$ identifies bonds whose retained state comes closest to exhausting its logical capacity.
This prescription preserves the norm optimality of Schmidt truncation and can accompany monitored Schmidt-truncation steps in DMRG, TEBD, and TDVP~\cite{White1992, Vidal2004, Haegeman2016, Paeckel2019}, including compact-MPO TEBD for Hamiltonians beyond nearest-neighbor interactions~\cite{Catalano2025TEBD}.
In each case, the certificate requires the normalized leading Schmidt spectrum and its associated discarded weight.
It bounds the resources of the state immediately before the monitored truncation.
It does not account for errors accumulated in earlier updates or provide error bounds for energies or local observables.


In conclusion, the occupied Schmidt support defines a common logical capacity for entanglement and nonlocal non-stabilizerness.
At the mathematical level, our central result is the joint constraint $(S_2)^2+(M_2^{\rm NLM})^2\leq q^2$.
For fixed logical capacity, the physically attainable resource pair $(S_2,M_2^{\rm NLM})$ lies within the quarter disk of radius $q$. 
This budget is nevertheless asymmetric: entanglement can exhaust the capacity, whereas the maximal Schmidt-representative non-stabilizerness satisfies $\mu(q)=3\log_2q+O(1)$ as $q\to\infty$ and therefore occupies an asymptotically vanishing fraction of it.
The uniform additive bound between the canonical and optimized resources implies that the supremum of $M_2^{\rm NLM}$ at fixed capacity also scales as $3\log_2q+O(1)$~\cite{SierantSpectral2026}.

From a many-body perspective, the logical Schmidt capacity is thus the finite-virtual-space budget shared by two resources that need not diagnose the same physics.
This distinction is relevant near quantum critical points and symmetry-protected phases, across ergodic-to-localized crossovers, and during nonequilibrium evolution, where nonlocal non-stabilizerness can reveal changes in the organization of the Schmidt spectrum that are not captured by entanglement alone~\cite{Catalano2025, Catalano2026, Liu2026, Li2026}.
Across every bipartition, our bound constrains the joint region accessible to a finite occupied Schmidt support.
For a single-edge tensor-network cut, this becomes a constraint set by the occupied bond dimension.
Long-range non-stabilizerness is instead defined through the obstruction to removing non-stabilizerness by shallow local circuits~\cite{Korbany2025}.
Comparing this circuit-based notion with the cut-resolved spectral constraint developed here, particularly at MPS renormalization-group fixed points, is a natural direction for future work.
The two notions are not equivalent: a shallow circuit may cross the bipartition and change its Schmidt spectrum, whereas the local-unitary transformations entering $M_2^{\rm NLM}$ preserve it.

The truncation certificate gives the discarded-weight threshold a quantitative interpretation in terms of resource bounds.
The retained spectrum and discarded weight provide a one-sided bound on the joint resources of the state immediately before each monitored Schmidt truncation.
Adjusting the retained rank controls the additive correction to this bound, while $\eta_\ell$ identifies cuts whose retained resources come closest to exhausting their logical capacity.
These diagnostics complement conventional checks on energies and local observables.
They do not, by themselves, certify convergence to the ground state or control errors accumulated during time evolution.

These results also complement existing MPS methods for evaluating stabilizer R\'enyi entropies of the full many-body state~\cite{HaugPiroli2023}, as well as numerical studies reporting rapid bond-dimension convergence for full-state and mutual non-stabilizerness in the spin-1 anisotropic Heisenberg chains considered in Ref.~\cite{Frau2024}.
By contrast, the resource studied here is defined from the Schmidt spectrum associated with a specified bipartition, making both the occupied-capacity constraint and the truncation certificate intrinsically cut-resolved.

Several questions remain open.
These include identifying the spectra that maximize the canonical and optimized resources at finite $q$ and determining the optimal additive terms in their large-capacity asymptotics.
The full region of attainable resource pairs, including the attainable portion of the circular boundary, also remains to be characterized.
Establishing whether the canonical representative always attains the local-unitary minimum would further clarify the relation between the two resources.
Resolving these questions would sharpen our understanding of which resource combinations a finite Schmidt support can sustain.

Acknowledgments.--- The author thanks Fabio Franchini for useful discussions and comments on the manuscript. The author acknowledges support from the project ``Implementation of cutting-edge research and its application as part of the Scientific Center of Excellence for Quantum and Complex Systems, and Representations of Lie Algebras," Grant No. PK.1.1.10.0004, co-financed by the European Union through the European Regional Development Fund under the Competitiveness and Cohesion Programme 2021--2027. The author also acknowledges support from the Croatian Science Foundation (HrZZ) through the project IP-2025-02-1667, ``Mining the Quantum: Frustration, Disorder, and Devices."


\clearpage
\onecolumngrid

\appendix

\section{Proof of the main bound}
\label{app:proof}

This Supplemental Material gives the analytical reductions and the finite
certification scheme for Eq.~(3) of the main text, and derives Eq.~(4).
All logarithms are to base two unless explicitly indicated otherwise.


\section{Notation and register formulation}

Let $\bm\lambda=(\lambda_0,\ldots,\lambda_{D-1})$ be a normalized, nonincreasing spectrum, padded with zeros to length $D=2^q$. 
Binary labels are identified with integers in their natural order. 
We use the notation of the main text,
\begin{align}
	P&=\sum_x\lambda_x^2, & S_2&=-\log_2 P,\\
	A_{u,k}&=\sum_x(-1)^{k\cdot x}\sqrt{\lambda_x\lambda_{x\oplus u}},
	&\Phi_q&=\frac1D\sum_{u,k}A_{u,k}^4,
	&\Msch&=-\log_2\Phi_q. \label{eq:definitions}
\end{align}
The sums run over $\F_2^q$, and binary additions below are modulo two.
We prove the stronger register statement
\begin{equation}
	S_2^2+(\Msch)^2\le q^2
	\label{eq:budget}
\end{equation}
for every ordered probability vector of length $2^q$, including those whose occupied support requires fewer than $q$ logical qubits. 
This formulation permits induction on registers whose spectra may contain zero entries.
Taking the minimal register size $q=\lceil\log_2 r_{\Sch}\rceil$ then gives Eq.~(3) of the main text. 
Appending a zero half leaves both $P$ and $\Phi_q$ unchanged, as also follows from the recursion derived below.
For $q=0$, the spectrum is $(1)$ and both entropies vanish.

The stabilizer purity $\Phi_q$ admits a useful representation in terms of the Gowers uniformity norm~\cite{Gowers2001, Tao2012}.
For a nonnegative function $f$ on $\F_2^q$, this norm is defined by
\begin{equation}
	\|f\|_{U^3}^8 = \E_{x,a,b,c}
	\prod_{\omega\in\{0,1\}^3}
	f(x+\omega_1a+\omega_2b+\omega_3c),
	\label{eq:gowers}
\end{equation}
where $\E$ denotes the uniform average over all four binary vectors $x,a,b,c$.
Taking $f(x)=\sqrt{D\lambda_x}$, Walsh orthogonality in Eq.~\eqref{eq:definitions} gives
\begin{equation}
	\Phi_q=\|f\|_{U^3}^8,
	\qquad
	\E f^2=1,
	\qquad
	\E f^4=DP.
\end{equation}
Indeed, expanding the fourth Walsh moment leaves only terms whose four indices sum to zero. 
These can be written as $x$, $x+a$, $x+b$, and $x+a+b$; the translation in $A_{u,k}$ provides the third direction of the cube in Eq.~\eqref{eq:gowers}.

We will use the standard monotonicity property
$|\E f|\le\|f\|_{U^2}\le\|f\|_{U^3}$, which follows from repeated Cauchy--Schwarz inequalities~\cite{Tao2012}.

For the recursive argument below, it is convenient to express
the same functional using unnormalized sums.
For a nonnegative vector $h$ on $\F_2^j$, with $N=2^j$, define
\begin{align}
	C_h(u,v)&=\sum_x h_xh_{x+u}h_{x+v}h_{x+u+v},\\
	Q(h)&=\sum_{u,v}C_h(u,v)^2.
	\label{eq:Q}
\end{align}
Expanding the square gives the cube sum, so that
$Q(h)=N^4\|h\|_{U^3}^8$. In particular,
\begin{equation}
	Q(\sqrt{\bm\lambda})=\Phi_q.
	\label{eq:QPhi}
\end{equation}
For two vectors $g$ and $h$ on the same register, we also define the mixed contribution
\begin{equation}
	T(g,h)=\sum_{u,v}C_g(u,v)C_h(u,v),
	\label{eq:QT}
\end{equation}
which enters the decomposition into two canonical halves.


\section{Endpoint estimates}

We first derive two bounds that are effective at opposite ends of the entanglement range.
Throughout this section, all $L^p$ norms and expectations are taken with respect to the uniform probability measure on $\F_2^q$.

Consider first the diagonal translation sector $u=0$. 
In this case,
\begin{equation}
	A_{0,k}=\sum_x(-1)^{k\cdot x}\lambda_x
\end{equation}
is the Walsh transform of the Schmidt spectrum itself. 
Parseval's identity gives
\begin{equation}
	\sum_k A_{0,k}^2 =D\sum_x\lambda_x^2=DP.
\end{equation}
Since this sector is one of the nonnegative contributions entering the stabilizer purity,
\begin{equation}
	\Phi_q =\frac1D\sum_{u,k}A_{u,k}^4
	\ge\frac1D\sum_k A_{0,k}^4.
\end{equation}
Applying Cauchy--Schwarz to the $D$ Walsh components yields
\begin{equation}
	\sum_k A_{0,k}^4 \ge\frac1D\left(\sum_k A_{0,k}^2\right)^2 =DP^2.
\end{equation}
We therefore obtain
\begin{equation}
	\Phi_q\ge P^2=2^{-2S_2},
	\qquad
	\Msch\le2S_2.
	\label{eq:lowendpoint}
\end{equation}
This bound proves the resource budget whenever
\begin{equation}
	S_2^2+(\Msch)^2 \le S_2^2+4S_2^2 =5S_2^2
	\le q^2,
\end{equation}
namely for $S_2\le q/\sqrt5$.

We next use the Gowers representation of the stabilizer purity.
For $f(x)=\sqrt{D\lambda_x}$, normalization and purity give
\begin{equation}
	\|f\|_2^2=\E f^2=1,
	\qquad
	\|f\|_4^4=\E f^4=DP.
\end{equation}
Interpolation between $L^1$ and $L^4$ gives
\begin{equation}
	\|f\|_2\le\|f\|_1^{1/3}\|f\|_4^{2/3}.
\end{equation}
Using $\|f\|_2=1$ and $\|f\|_4=(DP)^{1/4}$, we find
\begin{equation}
	\E f=\|f\|_1\ge(DP)^{-1/2}.
	\label{eq:meanlowerbound}
\end{equation}
The monotonicity of the Gowers norms,
$\|f\|_{U^1}=|\E f|\le\|f\|_{U^3}$, then gives
\begin{equation}
	\Phi_q=\|f\|_{U^3}^8
	\ge(\E f)^8
	\ge(DP)^{-4}
	=2^{-4(q-S_2)}.
\end{equation}
Equivalently,
\begin{equation}
	\Msch\le4(q-S_2).
	\label{eq:highendpoint}
\end{equation}
Hence
\begin{equation}
	S_2^2+(\Msch)^2\le S_2^2+16(q-S_2)^2=q^2+(S_2-q)(17S_2-15q).
\end{equation}
For $15q/17\le S_2\le q$, the last term is nonpositive, proving
the budget also in this high-entanglement region. The only interval not covered by the two endpoint estimates is therefore
\begin{equation}
	\frac q{\sqrt5}<S_2<\frac{15q}{17}.
	\label{eq:window}
\end{equation}


\section{Exact recursion and ordered mixed contribution}

Let $N=2^j$ and consider a nonnegative amplitude vector $F$ of length $2N$. 
We separate its first binary coordinate and write
\begin{equation}
	F(0,x)=g_x,\qquad F(1,x)=h_x,
	\qquad x\in\F_2^j.
\end{equation}
Thus $g$ and $h$ are the first and second halves of $F$, respectively. 

To derive the recursion for $Q$, recall from Eq.~\eqref{eq:Q} that it can be written as the following unnormalized cube sum:
\begin{equation}
	Q(F)= \sum_{z,a,b,c} \prod_{\omega\in\{0,1\}^3}
	F(z+\omega_1a+\omega_2b+\omega_3c),
\end{equation}
where $z,a,b,c\in\F_2^{j+1}$.
Let $\epsilon_0$ be the first bit of $z$, and let
$\epsilon_1,\epsilon_2,\epsilon_3$ be the first bits of the three directions $a,b,c$. 
The vertex indexed by $\omega$ then belongs to the half specified by
\begin{equation}
	\epsilon_0+\omega_1\epsilon_1
	+\omega_2\epsilon_2+\omega_3\epsilon_3
	\pmod 2.
\end{equation}
There are sixteen choices of these four bits.
If $\epsilon_1=\epsilon_2=\epsilon_3=0$, all eight vertices belong to the same half. 
The two possible values of $\epsilon_0$ therefore contribute $Q(g)$ and $Q(h)$.

For each of the seven nonzero choices of
$(\epsilon_1,\epsilon_2,\epsilon_3)$, exactly four vertices belong to each half. 
An invertible change of the three binary cube coordinates makes the half label depend on only one coordinate. 
The vertices in each half then form a square, and the two squares share the same two directions.
After summing over the remaining coordinates, their
contribution is 
\begin{equation}
	\sum_{x,u,v,d}
	\bigl[g_xg_{x+u}g_{x+v}g_{x+u+v}\bigr]
	\bigl[h_{x+d}h_{x+d+u}h_{x+d+v}h_{x+d+u+v}\bigr]
	=\sum_{u,v}C_g(u,v)C_h(u,v)
	=T(g,h).
\end{equation}
The equality follows by replacing $x+d$ with an independent summation variable. 
Both values of $\epsilon_0$ give the same contribution, since $T(g,h)=T(h,g)$.
The seven nonzero direction patterns therefore produce fourteen copies of $T(g,h)$, giving the exact identity
\begin{equation}
	Q(F)=Q(g)+Q(h)+14T(g,h).
	\label{eq:recursion}
\end{equation}

We next use canonical ordering to bound the mixed term from below. 
Suppose every entry of $g$ is at least every entry of $h$, and define
\begin{equation}
	H=\max_x h_x,\qquad b=\sum_xh_x^2.
\end{equation}
Here $b$ is the spectral weight of the second half when $g$ and $h$ are Schmidt-amplitude vectors.
Each of the $N$ products in $C_g(u,v)$ contains four entries no smaller than $H$. 
Hence
\begin{equation}
	C_g(u,v)\ge NH^4
\end{equation}
for every pair of translations $u,v$.
Since $h$ is nonnegative, $C_h(u,v)\ge0$, and therefore
\begin{equation}
	T(g,h)\ge NH^4\sum_{u,v}C_h(u,v).
\end{equation}

The remaining sum can be bounded using the unnormalized Walsh transform $\widehat h(k)=\sum_x (-1)^{k\cdot x}h_x$.
Expanding its fourth moment and applying Walsh
orthogonality gives
\begin{equation}
	\sum_{u,v}C_h(u,v)
	=\frac1N\sum_k\widehat h(k)^4
	\ge\frac1N\widehat h(0)^4
	=\frac1N\left(\sum_xh_x\right)^4.
\end{equation}
In the inequality we have retained only the zero Walsh mode; all fourth powers are nonnegative.
Combining the two estimates yields
\begin{equation}
	T(g,h)\ge\left(H\sum_xh_x\right)^4.
\end{equation}
Finally, $0\le h_x\le H$ implies $h_x^2\le Hh_x$.
Summing over $x$ gives $b\le H\sum_xh_x$, and hence
\begin{equation}
	T(g,h)\ge b^4.
	\label{eq:orderedmixed}
\end{equation}
This is the step where canonical ordering enters the lower bound.

A second estimate will be useful when the amplitude sums of the two halves are known. 
This estimate requires only nonnegativity, without any ordering assumption.
For each translation $d\in\F_2^j$, define
\begin{equation}
	a_d(x)=g_xh_{x+d},
	\qquad
	s_d=\sum_xa_d(x).
\end{equation}
Expanding $T(g,h)$ and writing the base point of the second square as $x+d$ gives
\begin{equation}
	T(g,h)=\sum_d\sum_{u,v}C_{a_d}(u,v).
\end{equation}
Applying the preceding Walsh fourth-moment bound to
each $a_d$ yields
\begin{equation}
	T(g,h)\ge\frac1N\sum_d s_d^4.
\end{equation}
Jensen's inequality for the $N$ translations gives
\begin{equation}
	\frac1N\sum_d s_d^4
	\ge\left(\frac1N\sum_d s_d\right)^4.
\end{equation}
Moreover, summing over all translations visits every pair of entries of $g$ and $h$ exactly once, so
\begin{equation}
	\sum_d s_d
	=\sum_{d,x}g_xh_{x+d}
	=\left(\sum_xg_x\right)\left(\sum_xh_x\right).
\end{equation}
We therefore obtain
\begin{equation}
	T(g,h)\ge
	\frac1{N^4}
	\left(\sum_xg_x\right)^4
	\left(\sum_xh_x\right)^4.
	\label{eq:mixedmean}
\end{equation}

To express these estimates compactly, set $X=N^{-1/2}\sum_xg_x$ and $Y=N^{-1/2}\sum_xh_x$.
The Gowers mean bound, with uniform averages on the $N$-element register, gives
\begin{equation}
	Q(g)=N^4\|g\|_{U^3}^8 \ge N^4\left(\frac1N\sum_xg_x\right)^8 =X^8,
\end{equation}
and similarly for $h$. 
Together with Eq.~\eqref{eq:mixedmean}, this gives
\begin{equation}
	Q(g)\ge X^8,\qquad
	Q(h)\ge Y^8,\qquad
	T(g,h)\ge X^4Y^4.
	\label{eq:meanbounds}
\end{equation}

We now apply the ordered recursion to the full Schmidt spectrum. 
Define the dyadic blocks and their weights as in the main text:
\begin{equation}
	\begin{gathered}
		B_0=\{0\},\qquad
		B_j=\{2^{j-1},\ldots,2^j-1\},
		\quad 1\le j\le q,\\
		W_j=\sum_{x\in B_j}\lambda_x.
	\end{gathered}
\end{equation}
These blocks partition the spectrum, so that $\sum_{j=0}^qW_j=1$.
Let
\begin{equation}
	h^{(j)}=(\sqrt{\lambda_0},\ldots,
	\sqrt{\lambda_{2^j-1}})
\end{equation}
be the amplitude vector of the first $2^j$ entries.
Its first half is $h^{(j-1)}$, while its second half contains precisely the block $B_j$ and has squared amplitude sum $W_j$.
Canonical ordering allows us to apply
Eq.~\eqref{eq:orderedmixed} at every step. The recursion then gives
\begin{equation}
	Q(h^{(j)}) \ge Q(h^{(j-1)})+14W_j^4,
\end{equation}
where we have dropped the nonnegative internal contribution of the second half.
Starting from $Q(h^{(0)})= (\sqrt{\lambda_0})^8 =W_0^4$ and iterating up to $j=q$, we obtain
\begin{equation}
	\Phi_q=Q(h^{(q)})
	\ge W_0^4+14\sum_{j=1}^qW_j^4.
	\label{eq:blockbound}
\end{equation}

It remains to bound this weighted fourth moment using only normalization. 
Set $c_0=1$ and $c_j=14$ for $j\ge1$. 
Writing each weight as $W_j=(c_j^{1/4}W_j) c_j^{-1/4}$ and applying H\"older's
inequality with conjugate exponents $4$ and $4/3$ gives 
\begin{equation}
	1=\sum_{j=0}^qW_j
	\le
	\left(\sum_{j=0}^qc_jW_j^4\right)^{1/4}
	\left(\sum_{j=0}^qc_j^{-1/3}\right)^{3/4}.
\end{equation}
Since $\sum_{j=0}^qc_j^{-1/3}=1+q/14^{1/3}$,
Eq.~\eqref{eq:blockbound} implies
\begin{equation}
	\Phi_q\ge\left(1+\frac q{14^{1/3}}\right)^{-3}.
\end{equation}
Taking the negative logarithm of the stabilizer purity finally yields
\begin{equation}
	\Msch\le3\log_2\left(1+\frac q{14^{1/3}}\right).
	\label{eq:logbound}
\end{equation}

The logarithmic bound above provides one of the two branches in Eq.~(4) of the main text.
The other follows by combining the quadratic resource budget, whose proof is completed below, with the bound $\Msch\le2S_2$, previously established in Refs.~\cite{Torre2026, Franchini2026} and recovered in 
Eq.~\eqref{eq:lowendpoint}.
Indeed, these inequalities give
$\frac54(\Msch)^2\le S_2^2+(\Msch)^2\le q^2$,
and hence $\Msch\le2q/\sqrt5$.
Taking the supremum over canonically ordered spectra at fixed logical Schmidt capacity therefore yields
\begin{equation}
	\mu(q)\le
	\min\left\{
	\frac{2q}{\sqrt5},
	3\log_2\left(1+\frac{q}{14^{1/3}}\right)
	\right\}.
\end{equation}
Thus, once the quadratic budget is established, Eq.~(4) of the main text follows.


\section{Analytical closure for $q\ge21$ and $q=1$}

Throughout the remaining window~\eqref{eq:window}, $\sqrt{q^2-S_2^2}\ge8q/17$. 
Thus Eq.~\eqref{eq:logbound} proves the budget whenever
\begin{equation}
	3\log_2\left(1+\frac q{14^{1/3}} \right) \le \frac{8q}{17}.
	\label{eq:largeq}
\end{equation}
For completeness, $14^{1/3}>12/5$. 
At $q=21$ it suffices to check $3\log_2(39/4) <168/17$, equivalent to the integer inequality
$39^{51}<2^{270}$. 
Moreover,
\begin{equation}
	\frac{d}{dq}\left[\frac{8q}{17}- 3\log_2\left(1+\frac{5q}{12}\right)\right]
	=\frac8{17} -\frac{15}{(12+5q)\ln2}>0\qquad(q\ge21).
\end{equation}
This establishes Eq.~\eqref{eq:budget} for all $q\ge21$.

For $q=1$, write $\bm\lambda=((1+z)/2,(1-z)/2)$, $0\le z\le1$.
Direct evaluation gives
\begin{equation}
	P=\frac{1+z^2}{2},\qquad \Phi_1=1-z^2+z^4.
\end{equation}
Since $(1-z^2+z^4)(1+z^2)=1+z^6\ge1$, we have
$\Msch\le\log_2(1+z^2)=1-S_2$. 
Therefore
$S_2^2+(\Msch)^2\le S_2^2+(1-S_2)^2\le1$.


\section{Purity reduction for $q=2,3,4$}

We now derive a lower bound on the stabilizer purity that depends only on the reduced-state purity $P$. 
This reduces the remaining verification to a one-variable inequality. 

For each translation $u\in\F_2^q$, define
\begin{equation}
	C_u=\sum_x\lambda_x\lambda_{x+u},
	\qquad
	B_u=\sum_x\sqrt{\lambda_x\lambda_{x+u}}.
\end{equation}
The correlators $A_{u,k}$ are the Walsh transform of $x\mapsto\sqrt{\lambda_x\lambda_{x+u}}$.
Parseval's identity therefore gives
\begin{equation}
	\sum_kA_{u,k}^2=DC_u.
\end{equation}
In particular, $C_0=P$ and $B_0=1$. Moreover,
\begin{equation}
	\sum_u C_u =\sum_{u,x}\lambda_x\lambda_{x+u} =\left(\sum_x\lambda_x\right)^2 =1,
\end{equation}
because, for each fixed $x$, the translations $x+u$ run over the entire register. 
Thus the total quadratic weight in the nonzero translation sectors is fixed by $1-P$.
Thus $\sum_{u\ne0}C_u=1-P$, or equivalently
$\sum_{u\ne0,k}A_{u,k}^2=D(1-P)$.

The binary translation structure also restricts which Walsh modes can contribute. 
Replacing $x$ with $x+u$ in the definition of $A_{u,k}$ gives
\begin{equation}
	A_{u,k}=(-1)^{k\cdot u}A_{u,k}.
\end{equation}
Hence $A_{u,k}=0$ whenever $k\cdot u=1$.
For $u\ne0$, exactly $D/2$ labels satisfy $k\cdot u=0$, so at most $D/2$ Walsh modes can be nonzero.
Among these, the zero mode is $A_{u,0}=B_u$.

We will also use two bounds relating $B_u$ and $C_u$.
Cauchy--Schwarz immediately gives $B_u^2\le DC_u$.
For the lower bound, note that a nonzero binary translation partitions the register into disjoint pairs $\{x,x+u\}$.
If $t_\ell=\sqrt{\lambda_x\lambda_{x+u}}$ denotes the contribution of the $\ell$th pair, then
\begin{equation}
	B_u=2\sum_\ell t_\ell,
	\qquad
	C_u=2\sum_\ell t_\ell^2.
\end{equation}
Since all $t_\ell$ are nonnegative, $B_u^2=4(\sum_\ell t_\ell)^2\ge4\sum_\ell t_\ell^2=2C_u$.
Consequently,
\begin{equation}
	2C_u\le B_u^2\le DC_u,
	\qquad u\ne0.
\end{equation}

We next bound the fourth moment separately in each translation sector. 
In the diagonal sector, $A_{0,0}=1$ and the remaining $D-1$ modes carry quadratic weight $DP-1$. 
Cauchy--Schwarz therefore yields
\begin{equation}
	\sum_kA_{0,k}^4
	\ge1+\frac{(DP-1)^2}{D-1}.
\end{equation}
For $u\ne0$, separating the zero Walsh mode leaves at most $D/2-1$ allowed modes, whose total quadratic weight is $DC_u-B_u^2$. 
Thus, for $D\ge4$,
\begin{equation}
	\sum_kA_{u,k}^4
	\ge B_u^4+\frac{(DC_u-B_u^2)^2}{D/2-1}.
\end{equation}
Summing these estimates over all translation sectors gives
\begin{equation}
	D\Phi_q\ge 1+\frac{(DP-1)^2}{D-1}
	+\sum_{u\ne0} \left[ B_u^4+\frac{(DC_u-B_u^2)^2}{D/2-1}
	\right]. \label{eq:sectorbound}
\end{equation}

To combine the nonzero sectors, we introduce
\begin{equation}
	Z=\sum_{u\ne0}B_u^2.
\end{equation}
There are $D-1$ such sectors. 
Applying Cauchy--Schwarz separately to their two contributions gives 
\begin{align}
	\sum_{u\ne0}B_u^4
	&\ge\frac{Z^2}{D-1},\\
	\sum_{u\ne0}(DC_u-B_u^2)^2
	&\ge\frac{[D(1-P)-Z]^2}{D-1},
\end{align}
where we used $\sum_{u\ne0}C_u=1-P$.
Equation~\eqref{eq:sectorbound} consequently becomes
\begin{equation}
	D\Phi_q\ge 1+\frac{(DP-1)^2}{D-1} +\frac1{D-1}
	\left[ Z^2+\frac{[D(1-P)-Z]^2}{D/2-1} \right].
	\label{eq:aggregateZ}
\end{equation}

We can constrain $Z$ in two complementary ways.
First, summing $B_u^2\ge2C_u$ over nonzero translations gives $Z\ge2(1-P)$.
Second, the identity
\begin{equation}
	\sum_uB_u =\sum_{u,x}\sqrt{\lambda_x\lambda_{x+u}} =\left(\sum_x\sqrt{\lambda_x}\right)^2
\end{equation}
and $B_0=1$ imply, again by Cauchy--Schwarz,
\begin{equation}
	Z\ge \frac{\left(\sum_{u\ne0}B_u\right)^2}{D-1}
	= \frac{\big[(\sum_x\sqrt{\lambda_x})^2-1\big]^2}{D-1}.
	\label{eq:Zmean}
\end{equation}
To turn this into a bound depending only on $P$, we therefore need the smallest possible value of $\sum_x\sqrt{\lambda_x}$ at fixed normalization and purity.

For $P<1$, this minimum is attained by a spectrum of the form 
\begin{equation}
	(\underbrace{a,\ldots,a}_{r-1\ \mathrm{entries}},
	b,0,\ldots,0),
	\qquad a\ge b\ge0.
\end{equation}
The reason for this structure can be seen by minimizing $R=\sum_x\sqrt{\lambda_x}$ on each face of the probability simplex. 
At a nonuniform stationary point, every positive entry satisfies a Lagrange-multiplier equation
\begin{equation}
	\frac1{2\sqrt{\lambda_x}}
	=\alpha+2\beta\lambda_x.
\end{equation}
The function $\ell(t)=1/(2\sqrt t)$ is strictly convex for $t>0$. 
Its intersection with an affine function has at most two points, so a stationary spectrum cannot have more than two distinct positive entries.

If these values are $a>b>0$, the smaller value can occur only once at a minimum. 
To see this, suppose two entries equal $b$ and vary them in opposite directions. 
This variation preserves both constraints to first order.
The corresponding second variation of the constrained objective has the sign of $\ell'(b)-2\beta$.
However, the stationary equations at $a$ and $b$ give
\begin{equation}
	2\beta=\frac{\ell(a)-\ell(b)}{a-b}>\ell'(b),
\end{equation}
where the strict inequality follows from convexity.
The second variation would therefore be negative,
excluding such a stationary point from being a minimum.
Thus all positive entries except possibly one must equal the larger value $a$. 
Uniform spectra are included as the limiting case $a=b$.

The normalization and purity constraints now reduce to
\begin{equation}
	(r-1)a+b=1,
	\qquad
	(r-1)a^2+b^2=P.
\end{equation}
Solving them gives
\begin{equation}
	\begin{gathered}
		r=\lceil1/P\rceil,
		\qquad
		\delta=\sqrt{\frac{rP-1}{r-1}},\\
		a=\frac{1+\delta}{r},
		\qquad
		b=\frac{1-(r-1)\delta}{r},\\
		R_{\min}(P)=(r-1)\sqrt a+\sqrt b.
	\end{gathered}
	\label{eq:Rmin}
\end{equation}
Indeed, the conditions $\delta\ge0$ and $b\ge0$ require $1/r\le P\le1/(r-1)$, which determines the support size away from the shared endpoints.
At those endpoints, the expressions from adjacent branches describe the same spectrum after zero padding.
For $P=1$, the spectrum is pure and $R_{\min}(1)=1$.

Combining the two lower bounds on $Z$, define
\begin{equation}
	Y(P)= \max\left\{ 2(1-P), \frac{[R_{\min}(P)^2-1]^2}{D-1} \right\}.
\end{equation}
Every spectrum of purity $P$ satisfies $Z\ge Y(P)$.
To substitute this lower bound into Eq.~\eqref{eq:aggregateZ}, we must check the direction of monotonicity.
For fixed $P$, write
\begin{equation}
	F_P(Z)=Z^2+\frac{[D(1-P)-Z]^2}{D/2-1}.
\end{equation}
Its derivative is
\begin{equation}
	F_P'(Z)
	=\frac{D}{D/2-1}\,[Z-2(1-P)].
\end{equation}
Thus $F_P$ is nondecreasing throughout the relevant range $Z\ge2(1-P)$. 
Since $Z\ge Y(P)\ge2(1-P)$, replacing $Z$ with $Y(P)$ preserves the lower bound.

We therefore define
\begin{equation}
	L_D(P)= 1+\frac{(DP-1)^2}{D-1}
	+\frac1{D-1} \left[ Y(P)^2+ \frac{[D(1-P)-Y(P)]^2}{D/2-1} \right]
	\label{eq:LD}
\end{equation}
and obtain the purity-dependent estimate
\begin{equation}
	\Phi_q\ge\frac{L_D(P)}D.
	\label{eq:puritycertificate}
\end{equation}

Since $\Msch=-\log_2\Phi_q$ and $S_2=-\log_2P$,
the quadratic resource budget follows if
\begin{equation}
	\frac{L_D(P)}D \ge 2^{-\sqrt{q^2-(-\log_2P)^2}}.
\end{equation}
The endpoint estimates already cover the complementary entropy regions. 
It therefore suffices to verify this one-variable inequality on
\begin{equation}
	2^{-15q/17}\le P\le2^{-q/\sqrt5},
\end{equation}
the purity interval corresponding to Eq.~\eqref{eq:window}.

For $q=2,3,4$, the exact interval verification uses respectively $4$, $6$, and $12$ subintervals. 
Their rational endpoints and certified margins are listed in Table~\ref{tab:new-purity}. 
Together with the endpoint estimates, these certificates establish the quadratic resource budget for these three values of $q$.


\section{Entropy-sensitive certificates for $8\le q\le20$}
Group the first two eigenvalues into a block of mass $w_0$ and size $n_0=2$. 
For $i=1,\ldots,q-1$, take the blocks $\{2^i,\ldots,2^{i+1}-1\}$ with masses $w_i$ and sizes $n_i=2^i$.
Set $c_0=3/4$ and $c_i=14$ for $i\ge1$.

For a normalized two-entry spectrum $(u,v)$, with $u+v=1$, setting $z=uv$ gives
\begin{equation}
	Q((\sqrt{u},\sqrt{v}))
	=1-4z+16z^2
	=\frac34+16\left(z-\frac18\right)^2
	\ge\frac34.
\end{equation}
Homogeneity, the ordered recursion, and blockwise Cauchy--Schwarz imply
\begin{equation}
	\Phi_q\ge\sum_{i=0}^{q-1}c_iw_i^4,\qquad
	\sum_iw_i=1,\qquad \sum_i\frac{w_i^2}{n_i}\le P.
	\label{eq:convexrelaxation}
\end{equation}
For any rational $\eta\ge0$ and rational $x_i\ge0$, define
\begin{equation}
	\nu=\min_i\left(4c_ix_i^3+\frac{2\eta x_i}{n_i}\right),\qquad A=\nu- \sum_i\left(3c_ix_i^4+\frac{\eta x_i^2}{n_i}\right).
\end{equation}
Convex tangents to $c_iw^4+\eta w^2/n_i$, summed over the blocks, prove the global lower bound
\begin{equation}
	\Phi_q\ge A-\eta P.
\end{equation}
Consequently, an entropy interval $S_2\in[L,U]$ is certified by
\begin{equation}
	A-\eta2^{-L}>2^{-\sqrt{q^2-U^2}}.
	\label{eq:dualcertificate}
\end{equation}
The witnesses need not be exact minimizers of the convex relaxation.
For $q=8,\ldots,20$, the verification uses a total of $78$ intervals with rational witnesses. 
The interval counts are reported in Table~\ref{tab:new-dual}.


\section{Boundary-sensitive induction for $q=5,6,7$}
Let $k=q-1$ and $n=2^k$. Split the ordered spectrum into two halves,
with masses $a$ and $b=1-a$, purities $p_0,p_1$, and common boundary
$\tau=\lambda_{n-1}$. Introduce the scaled variables
\begin{equation}
	\mathcal P=nP,\qquad v=np_0,\qquad u=np_1=\mathcal P-v,
	\qquad t=n\tau.
\end{equation}
Every ordered spectrum obeys
\begin{align}
	&\tfrac12\le a\le1,\qquad b\le t\le a,\notag\\
	&a^2\le v\le na^2-2(n-1)at+(n-1)t^2,\notag\\
	&b^2\le u\le\min\{bt,nb^2\},\qquad u+v=\mathcal P.
	\label{eq:feasibility}
\end{align}
The lower purity bounds follow from Cauchy--Schwarz. Writing each
first-half entry as $\tau+d_i$, with $d_i\ge0$ and
$\sum_i d_i=a-n\tau$, and using $\sum_i d_i^2\le(\sum_i d_i)^2$
gives the upper bound on $v$. The second-half cap $\lambda_x\le\tau$
gives $p_1\le b\tau$, while $p_1\le b^2$ follows from nonnegativity.
No positive lower bound on $\tau$ is imposed; reduced-rank spectra
and zero boundaries are included.

Let $g_i=\sqrt{\lambda_i}$, $h_i=\sqrt{\lambda_{n+i}}$, and
$X=n^{-1/2}\sum_i g_i$, $Y=n^{-1/2}\sum_i h_i$.
The bounds used in the relaxation are
\begin{align}
	X&\ge X_*:=\max\left\{\frac{a^{3/2}}{\sqrt v},\sqrt t,
	\frac{\sqrt{a-(n-1)t/n}}{\sqrt n}
	+\frac{n-1}{n}\sqrt t\right\},\notag\\
	Y&\ge Y_*:=\max\left\{\frac{b^{3/2}}{\sqrt u},\frac b{\sqrt t}\right\}.
	\label{eq:XY}
\end{align}
The first terms are the interpolation estimate already used for
Eq.~\eqref{eq:highendpoint}, now applied to each unnormalized half.
The floor gives $X\ge\sqrt t$.
For the other floor bound, repeatedly use
$\sqrt{\tau+d}+\sqrt{\tau+e}\ge\sqrt{\tau+d+e}+\sqrt\tau$:
the minimal amplitude sum concentrates the excess mass above the
floor into a single entry. The cap gives
$\sqrt{\lambda_x}\ge\lambda_x/\sqrt\tau$ in the second half.
If $b=0$, the second half is identically zero. We then set $Y_*=0$
and define both the second-half term and the mixed term in
Eq.~\eqref{eq:frontier} to vanish, without evaluating $s_1$ or any
expression of the form $0/0$.

Assume the register theorem at size $k$. The entropies of the
normalized halves are
\begin{equation}
	s_0=\log_2\frac{na^2}{v},\qquad
	s_1=\log_2\frac{nb^2}{u}\quad (b>0).
\end{equation}
Homogeneity and induction give
$Q(g)\ge a^4 2^{-\sqrt{k^2-s_0^2}}$ and, for $b>0$,
$Q(h)\ge b^4 2^{-\sqrt{k^2-s_1^2}}$.
For $b=0$, one has $Q(h)=0$, consistently with the convention above.
Combining these with Eqs.~\eqref{eq:orderedmixed},
\eqref{eq:meanbounds}, and \eqref{eq:recursion}, we obtain
\begin{align}
	\Phi_q\ge\mathcal R_q:={}&
	\max\{X_*^8,a^4 2^{-\sqrt{k^2-s_0^2}}\}
	+\max\{Y_*^8,b^4 2^{-\sqrt{k^2-s_1^2}}\}\notag\\
	&+14\max\{X_*^4Y_*^4,b^4\}.
	\label{eq:frontier}
\end{align}
It suffices to verify, on the enlarged feasible domain
of Eq.~\eqref{eq:feasibility},
\begin{equation}
	\mathcal R_q\ge
	2^{-\sqrt{q^2-[\log_2(n/\mathcal P)]^2}}
	\label{eq:frontiertarget}
\end{equation}
within the entropy window~\eqref{eq:window}.

The interval scheme starts from a rational outer enclosure of
\begin{equation}
	n2^{-15q/17}\le\mathcal P\le n2^{-q/\sqrt5},\qquad
	\tfrac12\le a\le1,\quad 0\le t\le1,\quad
	0\le v\le n2^{-q/\sqrt5}.
\end{equation}
Necessary constraints~\eqref{eq:feasibility} contract or discard boxes.
Each surviving box is accepted only if a directed lower bound on
$\mathcal R_q$ exceeds a directed upper bound on the target; otherwise
it is bisected into two covering children. For example,
$X\ge a_{\min}^{3/2}/\sqrt{v_{\max}}$ and
$Y\ge b_{\min}/\sqrt{t_{\max}}$.
In the floor term of Eq.~\eqref{eq:XY}, use $a_{\min},t_{\max}$
in the first square root and $t_{\min}$ in the second; a negative
lower enclosure of the first radicand is replaced by zero.
For the first inductive term, $v_{\max}/(na_{\min}^2)$ gives an
upper bound on normalized purity and hence a lower bound on $s_0$;
the mass prefactor is bounded below by $a_{\min}^4$.
Physical bounds $0\le s_0,s_1\le k$ are retained. If a half-mass
lower endpoint is zero, its inductive contribution can safely be
bounded by zero. The target increases as $\mathcal P$ decreases,
so its upper bound uses $\mathcal P_{\min}$.

The exact arithmetic, subdivision rules, and verification counts for these three cases are described in the Reproducible finite certificates section and summarized in Table~\ref{tab:new-frontier}.


\section{Reproducible finite certificates}

The accompanying ancillary package contains the exact certificate data
and a self-contained Python verifier using only the standard library.
Running \texttt{python -S verify\_all.py} checks every directed bound,
acceptance inequality, coverage condition, and traversal trace.

The accompanying ancillary package contains the exact certificate data
in \texttt{certificates.json} and a self-contained Python verifier,
\texttt{verify\_all.py}, using only the standard library.
Running \texttt{python -S verify\_all.py} from the same directory checks
interval coverage and every acceptance inequality, first for $q=2,3,4$,
then successively for $q=5,6,7$, and finally for
$q=8,\ldots,20$.
The dual witnesses were generated numerically and stored as exact
rationals; the supplied verifier only replays their exact checks.
No floating-point arithmetic enters any acceptance, exclusion,
contraction, or coverage decision.

All square roots of nonnegative rational numbers are enclosed at scale
$S=10^{12}$: if $x=p/d$, set $a=\lfloor\sqrt{\lfloor pS^2/d\rfloor}\rfloor$.
Then $a/S\le\sqrt{x}<(a+1)/S$, with equal endpoints used for an exact root.
Bounds on $2^{-j/64}$ have denominator $2^{48}$ and are obtained by taking
six successive integer square roots of $2^{3072-j}$.
Indeed, if the resulting integer is $z$, then
\begin{equation}
	z^{64}\le 2^{3072-j}<(z+1)^{64},\qquad
	\frac{z}{2^{48}}\le2^{-j/64}<\frac{z+1}{2^{48}}.
\end{equation}
Again the upper and lower bounds coincide in the exact case.
Comparisons with these bounds enclose $-\log_2 P$ on the grid $j/64$.
For $2^{-\sqrt{q^2-s^2}}$, an upper bound is obtained by rounding
the square root down to that grid; a lower bound uses upward rounding.
All strict comparisons are then comparisons of rational numbers.
For endpoint-coverage tests, rational bounds on natural logarithms are
obtained after range reduction to $1\le x\le2$ from the first $24$ terms
of
\begin{equation}
	\ln x=2\sum_{m\ge0}\frac{z^{2m+1}}{2m+1},
	\qquad z=\frac{x-1}{x+1},
\end{equation}
with the remaining tail bounded by a geometric series.

For $q=2,3,4$, the full list of intervals is given in
Table~\ref{tab:new-purity}. Each interval $[\ell,h]$ lies within a single
branch $1/r\le P\le1/(r-1)$ of $R_{\min}(P)$.
Here is an explicit lower enclosure used by the verifier.
First evaluate a lower bound $R_-$ on $R_{\min}(h)$ using directed square
roots. The function $R_{\min}(P)$ is nonincreasing on each branch,
since its derivative has the sign of $1/\sqrt a-1/\sqrt b\le0$.
Writing $[x]_+:=\max\{x,0\}$ for the positive part of $x$, set
$C=D(1-h)$, $d=D/2-1$, and
\begin{equation}
	y=\max\left\{2(1-h),\frac{[R_-^2-1]_+^2}{D-1}\right\},\qquad
	K(y,C)=\begin{cases}
		y^2+(C-y)^2/d,&y\le C,\\
		y^2,&y>C.
	\end{cases}
\end{equation}
The expression $K$ is the minimum of $Z^2+(C'-Z)^2/d$ over
$Z\ge y$ and $C'\ge C$: for $y\le C$ its minimum occurs at $Z=y$,
because $y\ge C/(d+1)=2(1-h)$; for $y>C$ take $C'=Z=y$.
Thus throughout the interval the sector estimate gives
\begin{equation}
	\Phi_q\ge B_-:=\frac1D\left[1+
	\frac{(D\ell-1)^2+K(y,C)}{D-1}\right].
\end{equation}
The target $2^{-\sqrt{q^2-(-\log_2P)^2}}$ is largest at $P=\ell$.
Let $T_+$ denote its upper enclosure obtained by the rules above.
The last column of Table~\ref{tab:new-purity} gives an integer $m$ such
that $B_--T_+\ge m\,10^{-6}>0$.
The union of the intervals in each group contains
$[2^{-15q/17},2^{-q/\sqrt5}]$; this coverage is also verified.

\begin{table}[htbp]\centering
	\caption{Complete purity certificates. All endpoints are exact rationals.
		The positive margin is at least $m\,10^{-6}$.}
	\label{tab:new-purity}
	\renewcommand{\arraystretch}{1.55}
	\begin{tabular}{ccccc}\toprule
		$q$ & $\ell$ & $h$ & $r$ & $m$\\\midrule
		2 & $\dfrac{29}{100}$ & $\dfrac{1}{3}$ & 4 & 3175 \\[4pt]
		2 & $\dfrac{1}{3}$ & $\dfrac{5}{12}$ & 3 & 51777 \\[4pt]
		2 & $\dfrac{5}{12}$ & $\dfrac{1}{2}$ & 3 & 111453 \\[4pt]
		2 & $\dfrac{1}{2}$ & $\dfrac{27}{50}$ & 2 & 170588 \\[4pt]
		\midrule
		3 & $\dfrac{159}{1000}$ & $\dfrac{1}{6}$ & 7 & 29718 \\[4pt]
		3 & $\dfrac{1}{6}$ & $\dfrac{11}{60}$ & 6 & 26079 \\[4pt]
		3 & $\dfrac{11}{60}$ & $\dfrac{1}{5}$ & 6 & 19129 \\[4pt]
		3 & $\dfrac{1}{5}$ & $\dfrac{1}{4}$ & 5 & 22452 \\[4pt]
		3 & $\dfrac{1}{4}$ & $\dfrac{1}{3}$ & 4 & 57327 \\[4pt]
		3 & $\dfrac{1}{3}$ & $\dfrac{79}{200}$ & 3 & 106193 \\[4pt]
		\midrule
		4 & $\dfrac{17}{200}$ & $\dfrac{1}{11}$ & 12 & 477 \\[4pt]
		4 & $\dfrac{1}{11}$ & $\dfrac{21}{220}$ & 11 & 18991 \\[4pt]
		4 & $\dfrac{21}{220}$ & $\dfrac{1}{10}$ & 11 & 1625 \\[4pt]
		4 & $\dfrac{1}{10}$ & $\dfrac{19}{180}$ & 10 & 9196 \\[4pt]
		4 & $\dfrac{19}{180}$ & $\dfrac{1}{9}$ & 10 & 182 \\[4pt]
		4 & $\dfrac{1}{9}$ & $\dfrac{17}{144}$ & 9 & 6592 \\[4pt]
		4 & $\dfrac{17}{144}$ & $\dfrac{1}{8}$ & 9 & 5840 \\[4pt]
		4 & $\dfrac{1}{8}$ & $\dfrac{1}{7}$ & 8 & 6506 \\[4pt]
		4 & $\dfrac{1}{7}$ & $\dfrac{1}{6}$ & 7 & 22682 \\[4pt]
		4 & $\dfrac{1}{6}$ & $\dfrac{1}{5}$ & 6 & 37085 \\[4pt]
		4 & $\dfrac{1}{5}$ & $\dfrac{1}{4}$ & 5 & 52554 \\[4pt]
		4 & $\dfrac{1}{4}$ & $\dfrac{29}{100}$ & 4 & 75905 \\[4pt]
		\bottomrule\end{tabular}\end{table}

For $q=8,\ldots,20$, the supplied certificate data contain the exact
interval endpoints and rational dual witnesses used by the verifier. For $8\le q\le20$, the corresponding block of \texttt{certificates.json} contains
the exact interval endpoints $L,U$, the nonnegative rational witness
$\eta$, and the $q$ nonnegative rational values $x_i$ for every interval.
The verifier recomputes $A$ from these witnesses and checks
\begin{equation}
	A-\eta\,\overline{2^{-L}}
	>\overline{2^{-\sqrt{q^2-U^2}}}.
\end{equation}
Throughout the finite certificates, $\overline{x}$ and $\underline{x}$ denote, respectively, rigorous rational upper and lower enclosures of $x$.
In the first term, $L$ is rounded down to the $1/64$ grid before evaluating the exponential.
The intervals exactly partition $[447q/1000,883q/1000]$, which contains the uncovered entropy window.
Table~\ref{tab:new-dual} gives the number of intervals and a lower bound on the smallest strict margin for each $q$. The full witness vectors are provided in the data file.

\begin{table}[htbp]\centering
	\caption{Dual certificates. The minimum acceptance margin is at least
		$m\,10^{-6}$. The total number of intervals is 78.}
	\label{tab:new-dual}
	\begin{tabular}{rrr}\toprule
		$q$ & Intervals & $m$\\\midrule
		8 & 25 & 352 \\
		9 & 9 & 1372 \\
		10 & 6 & 623 \\
		11 & 5 & 3124 \\
		12 & 4 & 317 \\
		13 & 4 & 483 \\
		14 & 4 & 538 \\
		15 & 3 & 468 \\
		16 & 3 & 284 \\
		17 & 3 & 141 \\
		18 & 4 & 1947 \\
		19 & 4 & 1746 \\
		20 & 4 & 1568 \\
		\bottomrule\end{tabular}\end{table}

For $q=5,6,7$, the verifier uses the boundary-sensitive relaxation
in $(\mathcal P,a,t,v)$ derived above. The initial purity range is
\begin{equation}
	n\,\underline{2^{-j_+/64}}\le\mathcal P
	\le n\,\overline{2^{-j_-/64}},\qquad
	j_+=\left\lceil64\frac{883q}{1000}\right\rceil,\quad
	j_-=\left\lfloor64\frac{447q}{1000}\right\rfloor,
\end{equation}
where $n=2^{q-1}$. 
Denoting the rational upper endpoint of this enclosure by $\mathcal P_{\max}:=n\,\overline{2^{-j_-/64}}$,
the remaining initial ranges are $1/2\le a\le1$, $0\le t\le1$, and $0\le v\le\mathcal P_{\max}$.
This is a rational outer enclosure of the entire intermediate window. 
Every contraction uses only necessary constraints of the relaxation.
Endpoint rounding in a contraction is outward, with denominator $10^{12}$.
Up to eight contraction passes are made per box. An unresolved box is bisected along its largest width relative to the corresponding initial width, with ties resolved by coordinate order $(\mathcal P,a,t,v)$.
Every accepted box satisfies a strict rational lower-bound versus upper-bound comparison. No sampling criterion is used.

The same data file contains the complete deterministic traversal trace
for each of $q=5,6,7$.
The trace records acceptance, infeasibility, or the bisection coordinate
at each node. The verifier reconstructs every node and checks the trace,
counts, and terminal decisions. Both children of every bisection are
processed, and every run ends with an empty stack.
The induction is applied successively after the $q=4$ certificate has
been verified. The counts are given in Table~\ref{tab:new-frontier}.

\begin{table}[htbp]\centering
	\caption{Exhaustive boundary-sensitive verification. All runs have zero
		pending boxes. The final column bounds the smallest margin among accepted
		boxes from below, in units of $10^{-12}$.}
	\label{tab:new-frontier}
	\begin{tabular}{rrrrrrr}\toprule
		$q$ & Processed & Accepted & Infeasible & Terminal & Depth & Margin\\\midrule
		5 & 2669 & 1333 & 2 & 1335 & 21 & 1737991 \\
		6 & 1507 & 754 & 0 & 754 & 19 & 8842759 \\
		7 & 1181 & 591 & 0 & 591 & 19 & 18458894 \\
		\bottomrule\end{tabular}\end{table}

All finite checks terminate successfully. Together with the analytical
cases $q=0,1$ and $q\ge21$, these certificates complete the quadratic
resource-budget proof. The bound on $\mu(q)$ then follows from that
budget, $M_2^{\rm Sch}\le2S_2$, and the ordered logarithmic bound.

\clearpage
\twocolumngrid

\end{document}